\documentclass[11pt]{article}
\usepackage{amssymb}
\usepackage{latexsym}
\usepackage{amsmath}
\usepackage{graphicx}
\usepackage{listings}
\usepackage{cite}

\def\al{\alpha}
\def\L{\Lambda}
\def\l{\lambda}
\def\k{\kappa}
\def\g{\gamma}
\title{\LARGE{Static spherically symmetric black hole in Einstein-power-Yang-Mills-dilaton theory and some aspects of its thermodynamics }}
\author{M. M. Stetsko\footnote{E-mail: mstetsko@gmail.com}\
\\
  {\small Department for Theoretical Physics, Ivan Franko National University of Lviv,}\\
{\small 12 Drahomanov Str., Lviv, UA-79005, Ukraine
         }}
\begin{document}
\maketitle

{\abstract{A static spherically symmetric solution is obtained and examined in the framework of Einstein-power-Yang-Mills-dilaton theory. To derive this exact solution a dilaton potential $V(\Phi)$ is taken into account. Thermodynamics of the black hole is also studied, namely we have obtained and investigated the temperature and heat capacity of the black hole which has allowed us characterize the stability of the black hole. Using the framework of the extended thermodynamic phase space we have derived the equation of state for the black hole. We have studied the Gibbs free energy of the black hole, which has the behaviour similar for other black holes with dilaton fields. To characterize the thermal behaviour of the black hole near the critical point we have written the Ehrenfest's equations and calculated Prigogine-Defay ratio.}}

\section{Introduction}
For the recent decades there is constant interest to the black holes with nonabelian fields. Nonabelian fields appear in the framework of String Theory and might be treated as the generalization of abelian fields. Historically firstly black holes in Einstein-Yang-Mills theory were considered in the works of Yasskin \cite{Yasskin_PRD75} and the following work of Kasuya \cite{Kasuya_PRD82}. Revival of interest to black holes with nonabelian field started in the late 80-ies of the last century \cite{Bartnik_PRL88,Bizon_PRL90,Volkov_JETP89}, namely the black holes were observed to be unstable in case of asymptotically flat geometry \cite{Straumann_PLB90}, but later black  holes' solutions were derived in anti-de Sitter case which were shown to be stable \cite{Torii_PRD95,Volkov_PRD96,Volkov_PRep98,Mavromatos_JMP98,
Winstanley_CQG99,Bjoraker_PRL00}. In the following decades there was growing interest to black holes with nonabelian fields not only in the framework of the standard General Relativity, but also in other gravitational frameworks, numerous solutions were obtained and studied \cite{Brihaye_PLB03,Kleihaus_PRD04,Radu_CQG05, Radu_PRD06,Mann_PRD06,Brihaye_PRD07,Manvelyan_PLB09,
Baxter_PRD07,Mazhari_PRD07,Mazhari_PRD08,Lerida_PRD09,Mazhari_PLB08,
Mazhari_GRG10,Mazhari_PRD11,Cvetic_PRD10,Ghosh_PLB11,Feng_PLB15,Kleihaus_CQG16}. It should be pointed out that in nonabelian case there are various gauge groups, but most of the works study the solutions with the gauge groups which are the most relevant to physics, namely the unitary and orthogonal groups. To obtain black holes' solutions it is necessary to choose an evident form for the gauge potential. One of the simplest forms which nevertheless allowed to derive interesting and important results is the so-called Wu-Yang ansatz \cite{Mazhari_PRD07,Mazhari_PRD08,Ghosh_PLB11}. We also note that the Wu-Yang ansatz gives rise to the so-called magnetic-type solutions.

Apart of the situations where the gauge field is minimally coupled to gravity there is deep interest to solutions where the gauge field might be coupled to other types of fields. Here we point out dilaton fields which are widely studied and as it is known they might have the String Theory origin, namely they naturally appear in low energy limit. They might be introduced by the dimensional reduction procedure and if there are additional gauge fields they would be coupled to the dilaton field in a non minimal way. Numerous black holes with dilaton field were obtained and examined \cite{Gibbons_NPB88, Garfinkle_PRD91,Witten_PRD91,Gregory_PRD93,Kallosh_PRD93,
Rakhmanov_PRD94,Poletti_PRD94,Chan_NPB95,Cai_PRD96,Gao_PRD04,
Yazadjiev_CQG05,Astefanesei_PRD06,Mann_JHEP06,Kunz_PLB06,
Brihaye_CQG07,Charmousis_PRD09,Sheykhi_PRD07,Fernando_PRD09,
Sheykhi_PRD14,Hendi_PRD15,Dehyadegari_PRD17,Pedraza_CQG19,
Bravo_PRD18,Goldstein_JHEP10, Arefeva_JHEP16,Stetsko_EPJC19}. But we also note that most part of the works where dilaton black holes are examined deals with the abelian gauge fields only, although the evident form of the gauge field lagrangians might be very different starting from ordinary form which gives rise to standard Maxwell's equations up to some sophisticated nonlinear Lagrangians which are of great interest nowadays. At the same time there is not a big number of works where black holes' solutions are obtained in case when the dilaton fields are coupled to some nonabelian gauge fields \cite{Kleihaus_PRD04,Radu_CQG05,Mann_PRD06,Mazhari_GRG10,
Feng_PLB15,Kleihaus_CQG16,Stetsko_EYM20,Stetsko_EMDYM20}. Most of these work focused on static solution, since the structure of the equations even in static case is quite complicated. Thus it is interesting to extend the class of solvable models with nonabelian fields. We also point out that the character of the solution substantially depends on the gauge group which is examined. We also point out that black holes' solutions were derived also for nonminimal coupling between gravity and nonabelian gauge fields \cite{Balakin_PRD16}.

Another line of research which is widely studied nowadays is related to the studies of the black holes with different types of nonlinear gauge fields. Among the very vast area of various nonlinear gauge field Lagrangians we point out the so called Born-Infeld type and the power-law nonlinearity. It is known that Born-Infeld model was introduced to derive finite fields at the origin and the finite self-energy. Nonlinear power law model was introduced to derive conformally invariant Lagrangian for the electromagnetic field in the dimensions of space-time different that four, although this model then was studied in broader context. For nonabelian field Born-Infeld and power-law types of Lagrangians were also examined \cite{Mazhari_PRD08,Mazhari_GRG10,Mazhari_PRD11}. It should be noted that in \cite{Mazhari_PRD11} the power-law nonlinear Yang-Mills  Lagrangian was considered in $F(R)$ gravity. Some other type of nonlinearity for Yang-Mills fields were studied  in \cite{Radu_PRD06,Brihaye_PRD07} where terms of higher order Yang-Mills hierarchy were taken into account.

In this work we consider Einstein-power-Yang-Mills-dilaton theory, namely the Yang-Mills Lagrangian is nonlinear and take power-law form. The gauge group is taken to be $SO(n)$ and for the gauge potential the Wu-Yang ansatz will be used. We point out that the linear Yang-Mills field were examined in \cite{Mazhari_GRG10} and our recent work \cite{Stetsko_EYM20}. We also note that in the paper \cite{Mazhari_GRG10} a solution of Bertotti-Robinson type was obtained which is close to an extreme black hole's solution, whereas in our work \cite{Stetsko_EYM20} we obtained the solution which is closer to non-extreme black hole. To obtain the exact solution we have taken into account the dilaton potential $V(\Phi)$, to derive the solution of the Bertotti-Robinson type the dilaton potential was not necessary. Here we follow the  line started in our previous work \cite{Stetsko_EYM20}, namely we also take into account the dilaton potential $V(\Phi)$, here we remark that the dilaton potential allows to take into consideration the cosmological constant. We also study thermodynamics of the black holes using the standard as well as extended phase space concept. In the framework of the extended thermodynamics we obtain and examine equation of state and the Gibbs free energy.

This paper is organized as follows. In the next section we obtain and examine static spherically symmetric black hole's solution. In the third section we investigate thermodynamics of the black holes, using the standard thermodynamic phase space, namely we obtain and study black hole's temperature and heat capacity and write the first law. In the fourth section we use extended thermodynamic phase space, namely we write and examine the equation of state, we obtain the extended first law and the Smarr relation. To investigate the thermal behaviour below the critical point in more details obtain and study the Gibbs free energy. In the fifth section we write the Ehrenfest's equations and calculate Prigogine-Defay ratio. Finally the sixth section contains some conclusions. 

\section{Static black hole in Einstein-power-Yang-Mills-dilaton theory}
We examine $n+1$--dimensional ($n\geqslant 3$) Einstein-power-Yang-Mills-dilaton theory supposing that the dilaton and Yang-Mills fields are coupled and their coupling to gravity is minimal. The action for the system can be written in the form:
\begin{eqnarray}\label{action_int}
 S=\frac{1}{16\pi}\int_{{\cal M}} {\rm d}^{n+1}x\sqrt{-g}\left(R-\frac{4}{n-1}\nabla^{\mu}\Phi\nabla_{\mu}\Phi-V(\Phi)-e^{-4\alpha\Phi/(n-1)}\left(Tr(F^{(a)}_{\mu\nu}F^{(a)\mu\nu})\right)^{p}\right)+\frac{1}{8\pi}\int_{\partial{\cal M}} d^{n}x\sqrt{-h}K, 
\end{eqnarray}
where $g$ is the determinant of the metric $g_{\mu\nu}$, $R$ is the scalar curvature, $\Phi$ is the dilaton field and $V(\Phi)$ is the dilaton potential, $F^{(a)}_{\mu\nu}$ is the gauge field tensor, $\al$ denotes the dilaton coupling constant. The second integral in the action (\ref{action_int}) is the Gibbons-Hawking-York (GHY) term which is introduced to make the variation problem well-defined. In the GHY term $h$ denotes the determinant of the boundary metric $h_{\mu\nu}$ and $K$ is the trace of the extrinsic curvature.

The gauge field tensor is defined as follows:
\begin{equation}\label{gauge_field}
F^{(a)}_{\mu\nu}=\partial_{\mu}A^{(a)}_{\nu}-\partial_{\nu}A^{(a)}_{\mu}+\frac{1}{2\bar{\sigma}}C^{(a)}_{(b)(c)}A^{(b)}_{\mu}A^{(c)}_{\nu},
\end{equation} 
and here $A^{(a)}$ is the gauge field potential, $C^{(a)}_{(b)(c)}$ are the structure constants for the corresponding gauge group and $\bar{\sigma}$ is the coupling constant for the nonabelian field. In this paper, as it has been mentioned above, we consider the gauge group to be $SO(n)$.

Variation of the action with respect to the metric $\delta g_{\mu\nu}$, the dilaton field $\delta\Phi$ and the gauge field potential $\delta A^{(a)}_{\mu}$  allows to obtain the equations of motion for the system which can written in the following form:
\begin{eqnarray}\label{einstein}
\nonumber R_{\mu\nu}=\frac{g_{\mu\nu}}{n-1}\left(V(\Phi)-(2p-1)e^{-4\al\Phi/(n-1)}\left(Tr(F^{(a)}_{\rho\sigma}F^{(a)\rho\sigma})\right)^p\right)+\\
\frac{4}{n-1}\partial_{\mu}\Phi\partial_{\nu}\Phi+2pe^{-4\al\Phi/(n-1)}
\left(Tr(F^{(a)}_{\rho\sigma}F^{(a)\rho\sigma})\right)^{p-1}Tr(F^{(a)}_{\mu\l}{F^{(a)\l}_{\nu}});
\end{eqnarray}
\begin{equation}\label{scal_eq}
\nabla_{\mu}\nabla^{\mu}\Phi=\frac{n-1}{8}\frac{\partial V}{\partial \Phi}-\frac{\al}{2}e^{-4\al\Phi/(n-1)}\left(Tr(F^{(a)}_{\rho\sigma}F^{(a)\rho\sigma})\right)^{p};
\end{equation}
\begin{equation}\label{YM_eq}
\nabla_{\mu}\left(e^{-4\al\Phi/(n-1)}\left(Tr(F^{(a)}_{\rho\sigma}F^{(a)\rho\sigma})\right)^{p-1}F^{(a)\mu\nu}\right)+\frac{1}{\bar{\sigma}}e^{-4\al\Phi/(n-1)}\left(Tr(F^{(a)}_{\rho\sigma}F^{(a)\rho\sigma})\right)^{p-1}C^{(a)}_{(b)(c)}A^{(b)}_{\mu}F^{(c)\mu\nu}=0.
\end{equation}
Here we examine a static spherically symmetric solution of the field equations and the metric can be chosen in the form:
\begin{equation}\label{metric}
 ds^2=-W(r)dt^2+\frac{dr^2}{W(r)}+r^2R^2(r)d\Omega^2_{n-1},
\end{equation}
where $d\Omega^2_{n-1}$ is the line element of $n-1$--dimensional unit hypersphere.

We utilize the so-called magnetic Wu-Yang ansatz to define the gauge field potential $A^{(a)}_{\mu}$, so the components  of the gauge potential can be represented in the form:
\begin{equation}\label{gauge_pot}
{\bf A}^{(a)}=\frac{q}{r^2}C^{(a)}_{(i)(j)}x^{i}dx^{j}, \quad r^2=\sum^{n}_{j=1}x^2_j,
\end{equation}
where indices $a$, $i$ and $j$ run the following ranges $2\leqslant j+1<i\leqslant n$ and $1\leqslant a\leqslant n(n-1)/2$ and here the new parameter $q$ is taken to be equal to the Yang-Mills coupling constant $\bar{\kappa}$ ($q=\bar{\kappa}$). The coordinates $x_i$ in the relation (\ref{gauge_pot}) can be represented by virtue of the following relation:
\begin{eqnarray}
\nonumber x_1=r\cos{\chi_{n-1}}\sin{\chi_{n-2}}\ldots\sin{\chi_1},\quad x_2=r\sin{\chi_{n-1}}\sin{\chi_{n-2}}\ldots\sin{\chi_1},\\
\nonumber  x_3=r\cos{\chi_{n-2}}\sin{\chi_{n-3}}\ldots\sin{\chi_1},\quad x_4=r\sin{\chi_{n-2}}\sin{\chi_{n-3}}\ldots\cos{\chi_1},\\
\nonumber \cdots \quad\quad\\
x_n=r\cos{\chi_1}
\end{eqnarray}
The angular variabels $\chi_1,\ldots,\chi_{n-1}$ also allow to define the line element of the unit hypersphere in metric (\ref{metric}), namely we write:
\begin{equation}
d\Omega^2_{n-1}=d\chi^2_{1}+\sum^{n-1}_{j=2}\prod^{j-1}_{i=1}\sin^2{\chi_{i}}d\chi^2_{j},
\end{equation}
and the angular variables vary in the ranges $0\leqslant\chi_{i}\leqslant\pi, i=1,\ldots,n-2$, $0\leqslant\chi_{n-1}\leqslant 2\pi$.

Using the Wu-Yang ansatz for the gauge potential (\ref{gauge_pot}) one can easily calculate the gauge field tensor (\ref{gauge_field}) and then  derive the Yang-Mills field invariant, which takes the same form as for the linear gauge field case \cite{Stetsko_EYM20}:
\begin{equation}
Tr(F^{(a)}_{\rho\sigma}F^{(a)\rho\sigma})=(n-1)(n-2)\frac{q^2}{r^4R^4}.
\end{equation}
It is easy to check that the equations of motion for the gauge field (\ref{YM_eq}) are satisfied.

\begin{figure}
\centerline{\includegraphics[scale=0.3,clip]{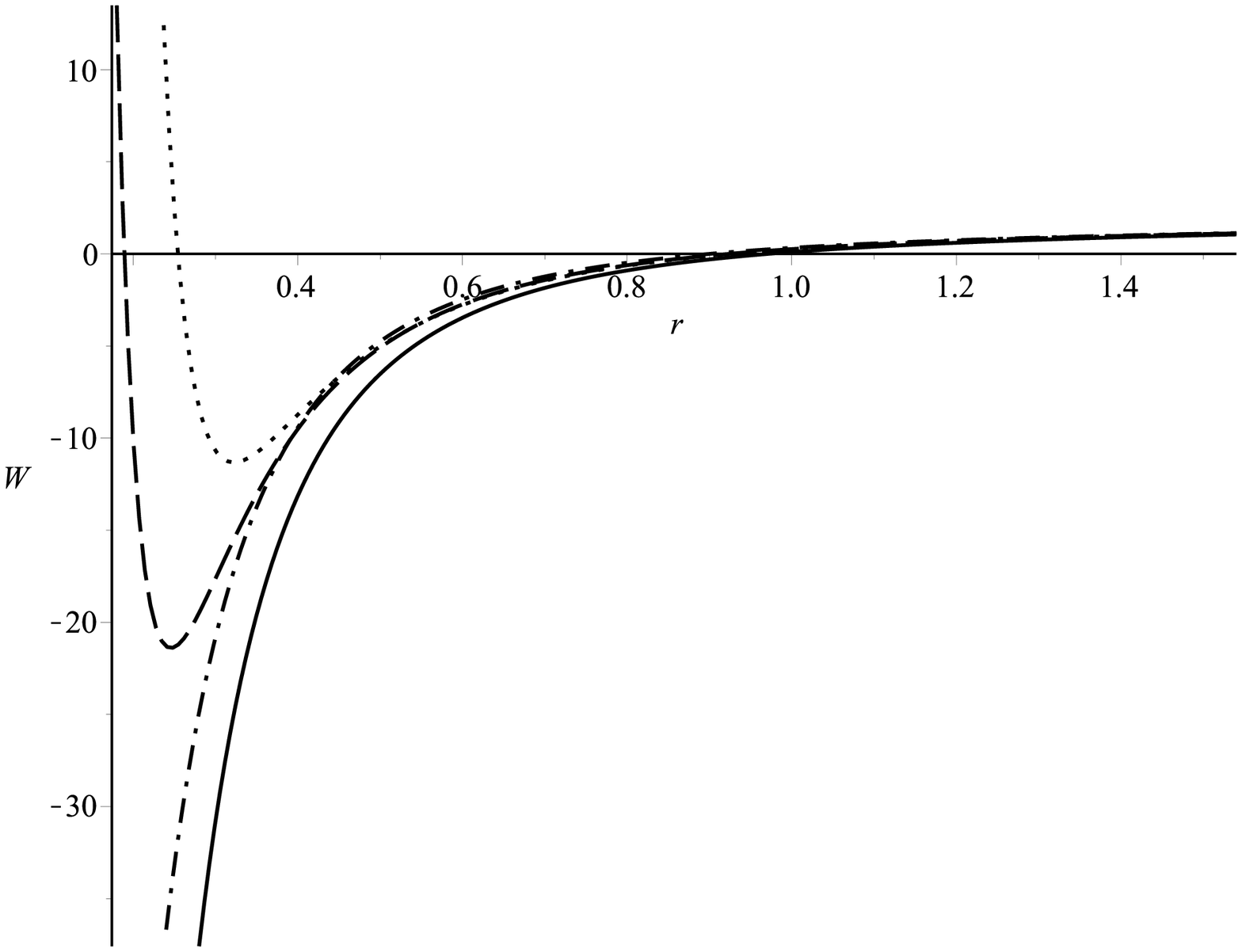}\includegraphics[scale=0.3,clip]{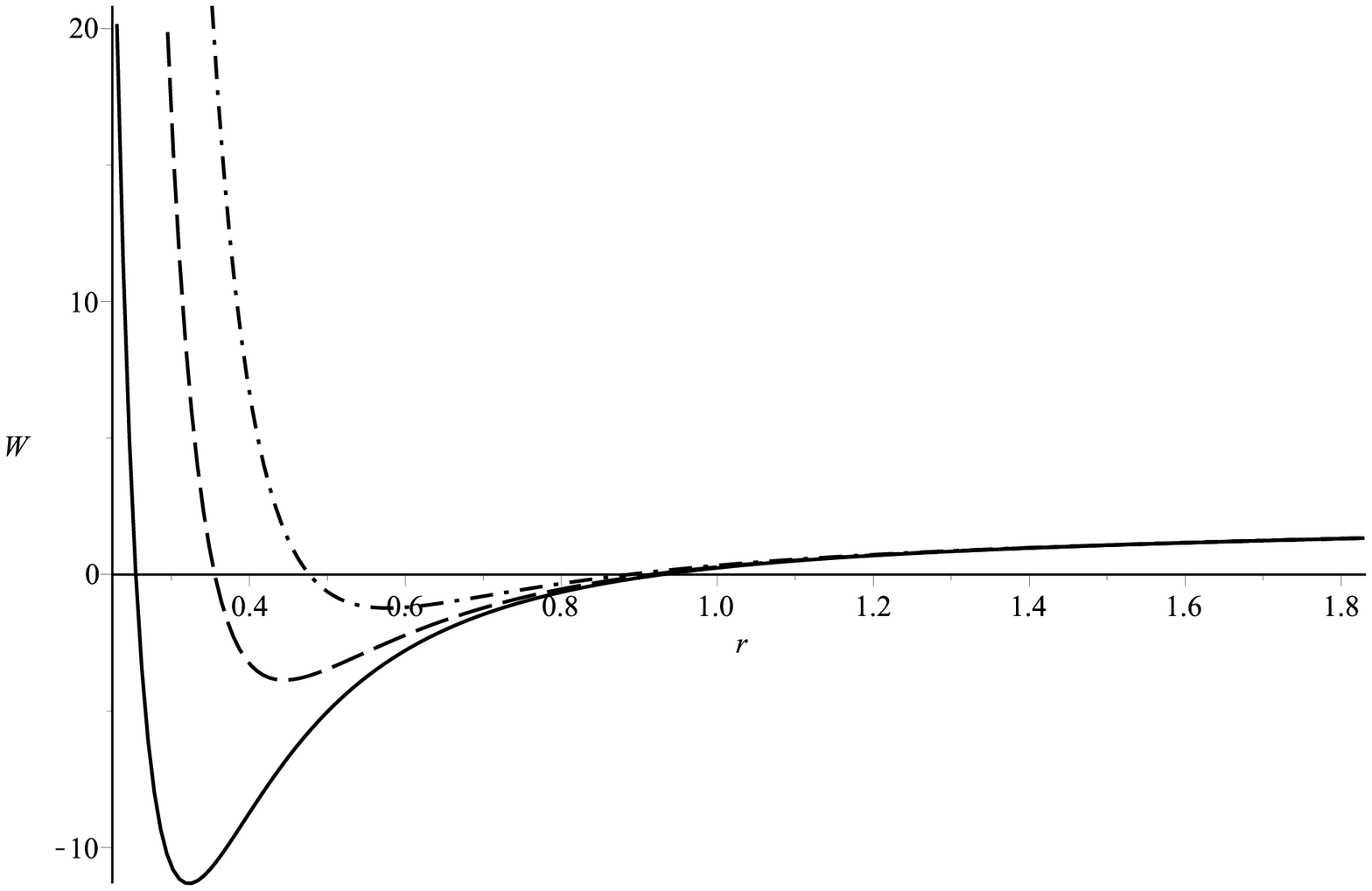}}
\caption{Metric function W(r) for various values of the parameter $p$ (the left graph) and the gauge field parameter (gauge charge) $q$ (the right one). For the both graphs it is chosen: $n=5, b=1, m=1,\al=0.2, \L=-2$. For the left graph the parameter $p$ takes the following values: $p=1$, $p=1.5$, $p=1.8$ and $p=2$ for solid, dash-dotted, dashed and dotted curves respectively and $q=0.2$. For the right graph the  parameter $q$ takes the values $q=0.2$, $q=0.25$ and $q=0.3$ for solid, dash-dotted and dashed curves correspondingly, while $p=2$.}\label{metr_f_graph}
\end{figure}

Similarly as it was done for the linear gauge field case \cite{Stetsko_EYM20} here we choose the potential for the dilaton field $V(\Phi)$ in the Liouville-type form, namely we take:
\begin{equation}\label{dilat_pot}
V(\Phi)=\Lambda e^{\lambda\Phi}+\Lambda_1 e^{\lambda_1\Phi}+\Lambda_2e^{\lambda_2\Phi}.
\end{equation}
The parameters of the potentials, namely $\l$, $\l_i$ and $\L_i$ where $i=1,2$ are chosen to fulfill the equations of motion for the gravitational as well as for the dilaton fields. It should be noted that the Liouville-type potentials are used in numerous papers where dilaton fields were studied.  We also take the function $R(r)$ in the following form:
\begin{equation}\label{ansatz_R}
R(r)=e^{2\al\Phi/(n-1)}.
\end{equation}
This ansatz and the evident form of the dilaton potential (\ref{dilat_pot}) allow us to obtain the solutions of the equations of motion (\ref{einstein}) and (\ref{scal_eq}). As a result we can write:
\begin{eqnarray}
\nonumber W(r)=-mr^{1+(1-n)(1-\gamma)}+\frac{(n-2)(1+\al^2)^2}{(1-\al^2)(\al^2+n-2)}b^{-2\gamma}r^{2\gamma}-\\\frac{\Lambda(1+\al^2)^2}{(n-1)(n-\al^2)}b^{2\gamma}r^{2(1-\gamma)}+\frac{p(n-1)^{p-1}(n-2)^{p}(1+\al^2)^2q^{2p}}{(\al^2-p)(n+3\al^2-4p)}b^{-2(2p+1)\gamma}r^{2(1+\gamma)+4p(\g-1)}\label{W_nln}
\end{eqnarray}
and here $\gamma=\al^2/(1+\al^2)$.
\begin{equation}
\Phi(r)=\frac{\al(n-1)}{2(1+\al^2)}\ln{\left(\frac{b}{r}\right)}.
\end{equation}
We point out that in the metric function (\ref{W_nln}) the parameter $m$ is an integration constant related to the mass of the black hole. The parameter $b$ is the other integration constant which appears due to integration of the equations of motion for the dilaton field $\Phi$, but the physical meaning of this constant is not so transparent as for the constant $m$. The parameters of the Liouville potential take the form as follows:
\begin{gather}
\l=\frac{4\al}{n-1},\quad\l_1=\frac{4}{\al(n-1)},\quad \l_2=\frac{4(2p-\al^2)}{\al(n-1)};\\
\L_1=\frac{(n-1)(n-2)\al^2}{\al^2-1}b^{-2},\quad \L_2=\frac{\al^2}{p-\al^2}(n-1)^p(n-2)^pq^{2p}b^{-4p}.
\end{gather}
In the limit $p=1$ the relations for $\l_2$ and $\L_2$ give rise to corresponding relations for linear field case, derived in our previous work \cite{Stetsko_EYM20}. It should be noted that the metric function $W(r)$ is ill-defined when $\al=1$, this peculiarity is known as a string singularity and when $\al=\sqrt{n}$, similar behaviour of the metric function takes place for other types of the dilaton black holes. In the limit $\al\to 0$ the metric function $W(r)$ can be represented in the form:
\begin{equation}
W(r)=1-\frac{m}{r^{n-2}}+\frac{\L}{n(n-1)}r^2-\frac{(n-1)^{p-1}(n-2)^{p}q^{2p}}{(n-4p)}r^{2(1-2p)}.
\end{equation}
For nonzero dilaton parameter ($\al\neq 0$) the metric function $W(r)$ (\ref{W_nln}) is rather complicated, but nonetheless some important features of this function can be described using its evident form (\ref{W_nln}). Firstly, for very small values of $r$ the dominant terms in the metric function might be of the Schwazschild type $\sim -mr^{1+(1-n)(1-\g)}$ or caused by the gauge field $\sim q^{2p}r^{2(1+\g)+4p(\g-1)}$ and the first of them is always negative, whereas the second one might be positive or negative depending on the relation between the parameters $\al$ and $p$. If the parameter of nonlinearity $p>1$ but it is close to one, the dominant term at small distances would be of the Schwarzschild type, the same conclusion takes place also when $p<1$. In case of increasing of the parameter of nonlinearity the leading term near the origin is that one which is caused by the gauge field, namely it takes place when $p>(n-1+(1+2\g)/(1-\g))/4$ and this fact is illustrated  on the left graph of the Figure [\ref{metr_f_graph}].  When the gauge field term in the metric function $W(r)$ becomes dominant, the black hole might have two horizons, similarly as it is for the Einstein-Maxwell-Yang-Mills-dilaton black hole \cite{Stetsko_EMDYM20}, but the increase of the parameter $q$ makes these horizons closer (the right graph on the Figure [\ref{metr_f_graph}]) with their following merging (extreme case) and then the horizon disappears, so our solution gets transformed into a naked singularity. We note, that similar behaviour takes place even for the standard Reissner-Nordstrom black holes when one increases the electric charge if the mass of the black hole is held fixed. For large distances the metric is not asymptotically flat, but it is not exactly of AdS- or dS-types, it becomes of these types when $\al\to 0$, actually the behaviour at large distances ($r\rightarrow\infty$) is typical for the black holes with dilaton field and and it is almost the same as for the Einstein-Yang-Mills-dilaton black hole with the linear Yang-Mills field \cite{Stetsko_EYM20}.

To understand the behaviour of the solution (\ref{W_nln}) better we also examine the Kretschmann scalar $R_{\k\l\mu\nu}R^{\k\l\mu\nu}$ which allows one to verify what points are the points of true physical singularity. It can be verified that at the horizons the Krestchmann scalar is not singular and this point is the point of a coordinate singularity as it always happens for a black hole. The only point of the physical singularity is the origin and the character of behaviour of the Kretschmann scalar at this point is defined by the leading term of the metric near the origin. If the leading term is of the Schwarzschild type then we have: $R_{\k\l\mu\nu}R^{\k\l\mu\nu}\sim m^2r^{2(1-n)(1-\g)-2}$, whereas in the case when the dominant term is caused by the gauge field we arrive at: $R_{\k\l\mu\nu}R^{\k\l\mu\nu}\sim q^{4p}r^{4(\g-2p(\g-1))}$, but in both cases the Kretschmann scalar is divergent, what confirms the fact that the origin is the point of physical singularity.

\section{Black hole's thermodynamics}
\begin{figure}
\centerline{\includegraphics[scale=0.33,clip]{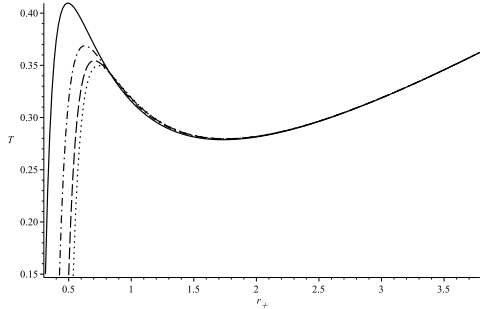}}
\caption{Black hole's temperature $T$ as a function of horizon radius  $r_+$ for various values of the parameter $p$. The correspondence of the lines is as follows: the solid, dash-dotted, dashed and dotted lines correspond to the $p=1.2, p=1.5, p=1.8, p=2$ respectively. The other parameters are fixed for all the curves and take the following values: $n=5, \al=0.1, b=1, \L=-4, q=0.2$.}\label{temp_graph}
\end{figure}

We start the investigation of the thermodynamics of the black hole calculating its temperature. The temperature is defined in a standard way, namely we use the surface gravity concept, so we can write:
\begin{eqnarray}\label{temp_nln}
T=\frac{W'(r_+)}{4\pi}=\frac{(1+\al^2)}{4\pi}\left(\frac{n-2}{1-\al^2}b^{-2\gamma}r^{2\gamma-1}_{+}-\frac{\Lambda}{n-1}b^{2\gamma}r_+^{1-2\gamma}+\frac{p(n-1)^{p-1}(n-2)^p}{\al^2-p}q^{2p}b^{-2(2p+1)\gamma}r_+^{1+2\gamma+4p(\g-1)}\right), 
\end{eqnarray}
and here $r_+$ is the event horizon of the black hole. The expression we have written for the temperature (\ref{temp_nln}) allows us to make some general conclusions about the behaviour of this function. Before we make these conclusions, we suppose that $\L$ is negative, actually in some sense it is analogous to the examination of the AdS-case, this assumption is made to avoid the existence of the cosmological horizon. We also assume that $\al<1$, the case with $\al>1$ will be considered elsewhere. Taking into account these assumptions we can conclude that for large values of the horizon's radius $r_+$ the behaviour of the temperature is mainly defined by the term $\sim\L r^{1-2\g}_+$, and we see that it increases with increasing of $r_+$ and this increase becomes slower when the dilaton parameter $\al$ goes up and if the dilaton parameter $\al\rightarrow 0$ the behaviour of this term becomes almost linear as it should be for an AdS black hole. For small radius of the horizon the temperature is mainly defined by the other two terms in the expression (\ref{temp_nln}) and if the gauge field term in the metric function $W(r)$ becomes the leading one the corresponding term also makes dominant contribution into the temperature. For intermediate values of the radius $r_+$ the temperature has nonmonotonous behaviour what is reflected on the Figure [\ref{temp_graph}]. From the Figure [\ref{temp_graph}] we see that the peak is higher for lower parameter of nonlinearity $p$ when the other parameters are held fixed. It should be also pointed out that the variation of the parameter $\al$ affects on the temperature for all the range of the radius of the horizon $r_+$ and the variation of the parameter $\L$ becomes substantial just for large radius of the horizon and these conclusions are completely the same as for the black hole with the linear Yang-Mills field \cite{Stetsko_EYM20}.  

\begin{figure}
\centerline{\includegraphics[scale=0.33,clip]{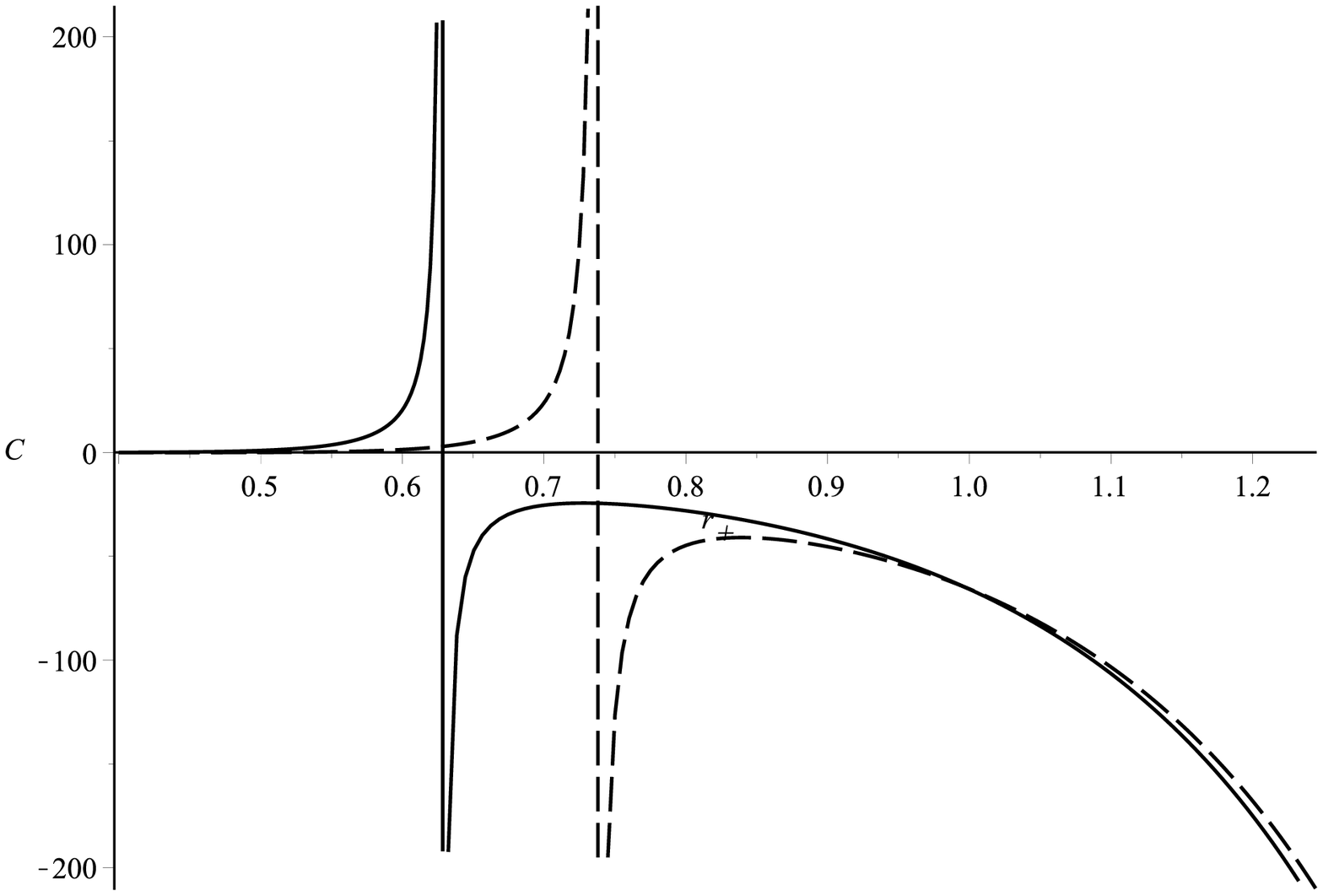}\includegraphics[scale=0.33,clip]{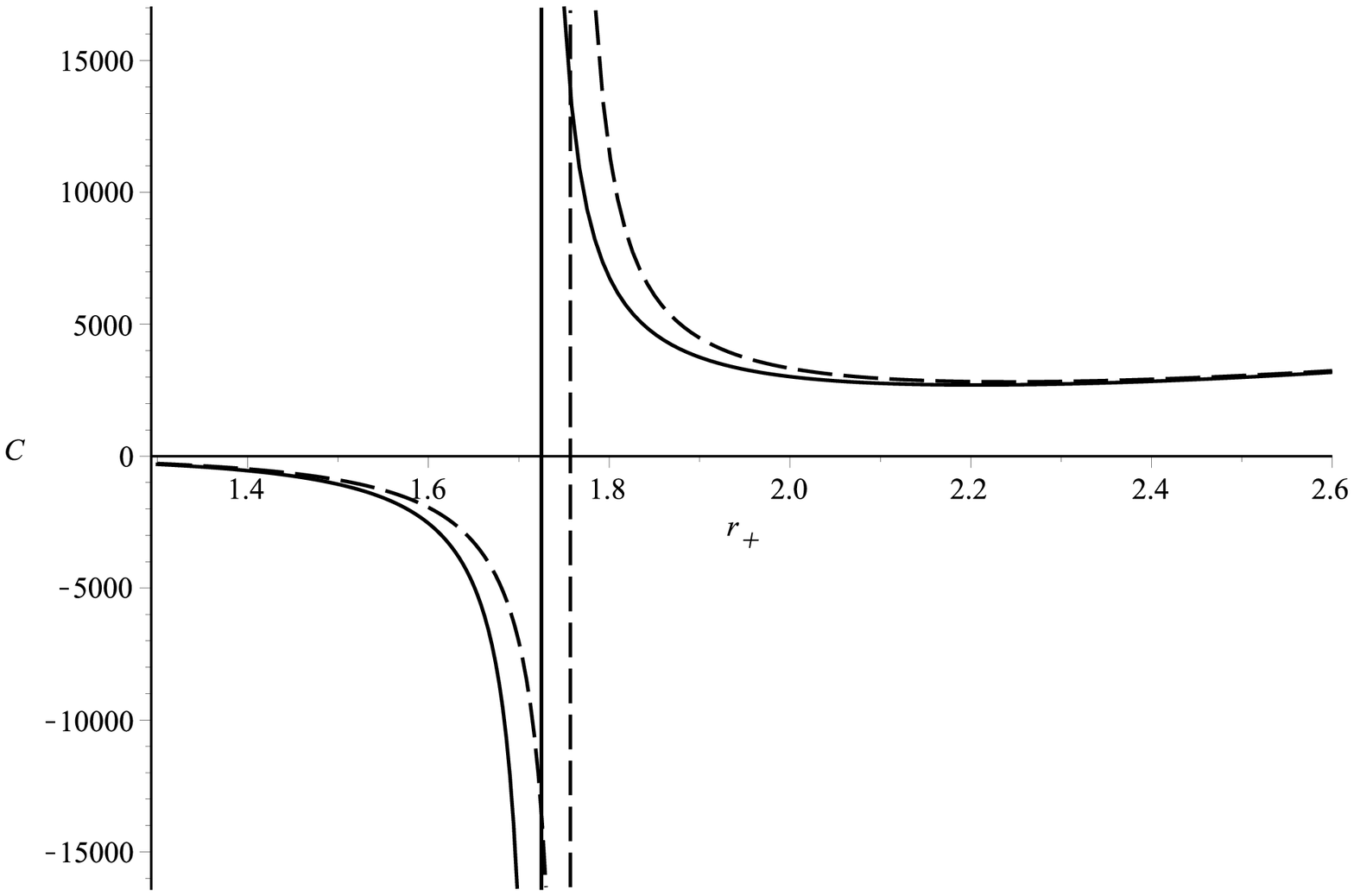}}
\caption{Heat capacity of the black hole $C_{\L,q}$ as a function of the horizon radius  $r_+$  with two discontinuity points. The fixed parameters are as follows: $n=5$, $\al=0.1$, $\L=-4$, $b=1$ and $q=0.2$. The correspondence of the  curves is as follows:  on the left graph the solid and the dashed lines correspond to $p=1.5$ and $p=2$ respectively, whereas on the right graph we have $p=1$ and $p=2$ again for the solid and the dashed lines.}\label{heatcap_gr}
\end{figure}

Now we can define the entropy of the black hole and since from the gravity side of view we have standard Einstein-Hilbert action and the fields are minimally coupled to gravity the entropy of the black hole will be a quarter of the horizon area, so we write:
\begin{equation}\label{entropy}
S=\frac{\omega_{n-1}}{4}b^{(n-1)\g}r^{(n-1)(1-\gamma)}_+,
\end{equation}
where $\omega_{n-1}$ is the surface area of a $n-1$--dimensional unit hypersphere. Since the other parameters of the metric (\ref{W_nln}) are held fixed, thus we can write the first law of black hole's thermodynamics:
\begin{equation}\label{first_law}
dM=TdS,
\end{equation}
and here the temperature should be treated as a thermodynamic entity:
\begin{equation}\label{temp_TD}
T=\left(\frac{\partial M}{\partial S}\right)_{\L,q}.
\end{equation}
The latter two relations allow us to derive thermodynamic mass of the back hole which takes the following form: 
\begin{equation}\label{mass_bh}
M=\frac{(n-1)b^{(n-1)\gamma}\omega_{n-1}}{16\pi(1+\al^2)}m.
\end{equation}
Since we examine the thermodynamics of the black holes the important issue here is the thermal stability, which can investigated by virtue of the heat capacity. The heat capacity is defined as follows:
\begin{equation}\label{heat_cap}
C_{\L,q}=T\left(\frac{\partial S}{\partial r_+}\right)_{\L,q}\left(\frac{\partial T}{\partial r_+}\right)^{-1}_{\L,q}.
\end{equation}
Having used the relations for the temperature (\ref{temp_nln}) and the entropy (\ref{entropy}) we write:
\begin{eqnarray}
\nonumber C_{\L,q}=\frac{(n-1)\omega_{n-1}}{4}b^{(n-1)\g}r^{(n-1)(1-\gamma)}_+\left(\frac{n-2}{1-\al^2}b^{-2\gamma}r^{2\gamma-1}_{+}-\frac{\Lambda}{n-1}b^{2\gamma}r_+^{1-2\gamma}+\right.\\\left.\nonumber\frac{p(n-1)^{p-1}(n-2)^p}{\al^2-p}q^{2p}b^{-2(2p+1)\gamma}r_+^{1+2\gamma+4p(\g-1)}\right)\left(-(n-2)b^{-2\gamma}r^{2\gamma-1}_{+}-\right.\\\left.\frac{\Lambda(1-\alpha^2)}{n-1}b^{2\gamma}r_+^{1-2\gamma}+\frac{p(n-1)^{p-1}(n-2)^p(1+3\al^2-4p)}{\al^2-p}q^{2p}b^{-2(2p+1)\gamma}r_+^{2\gamma+4p(\g-1)}\right)^{-1}.\label{heat_capac}
\end{eqnarray} 
Since the temperature (\ref{temp_nln}) might have nonmonotonous behaviour it gives rise to the conclusion that the heat capacity (\ref{heat_capac}) would be a discontinuous function and the points of the discontinuity are exactly the points where the temperature takes extreme values and it is reflected on the Figure [\ref{heatcap_gr}] where for convenience these points are shown separately. From the Figure [\ref{heatcap_gr}] we might conclude that the increase of the parameter $p$ does not affect seriously on qualitative features of the heat capacity $C_{\L,q}$, it only shifts these discontinuities to the right and for the left point of the discontinuity the shift is more substantial than for the right one. The Figure [\ref{heatcap_gr}] allows us to infer that the behaviour of the heat capacity we have derived here is typical for other types for the black holes with dilaton fields \cite{Stetsko_EYM20,Stetsko_EMDYM20,Stetsko_EPJC19}. Namely, for large values of the horizon radius $r_+$ we have positive heat capacity and the black holy is thermally stable. It should be pointed out that in the asymptotic domain of extremely large values of the horizon radius ($r_+\to\infty$) the heat capacity increases when the radius $r_+$ goes up and this is very similar to what we have for the black holes with asymptotically non-flat background, in particular for black holes with AdS asymptotic. Then for some intermediate values of the radius $r_+$ the heat capacity is negative and here we have the domain of thermally unstable black hole. Finally, for smaller values of $r_+$ we initially have positive values of the heat capacity. We also note that the increase of the parameter $q$ similarly to the variation of the parameter $p$ shifts these discontinuities to the right and when the absolute value of the cosmological constant $\L$ goes up these discontinuity points become closer and finally they merge. After their merging the heat capacity becomes positive in all this domain. This fact has simple explanations, the increase of the module of $\L$ gives rise to the consequence that the dependence $T=T(r_+)$ becomes monotonically increasing function, thus the heat capacity $C_{\L,q}$ would not have any discontinuity and it also takes positive values almost everywhere on its domain.
\section{Extended phase space thermodynamics}
Extended phase space thermodynamics is an area of active investigation  in recent decade. The extension of phase space allowed to develop relations between standard condensed matter thermodynamics and black hole thermodynamics in full generality \cite{Kubiznak_CQG17}. To obtain the extended phase space it is assumed that the cosmological constant $\L$ might be varied and as a thermodynamic quantity it is directly related to thermodynamic pressure \cite{Kastor_CQG09,Cvetic_PRD11,Dolan_CQG11}. Here we introduce the pressure similarly as it was performed for other dilaton black holes, namely we write:
\begin{equation}\label{press}
P=-\frac{\Lambda}{16\pi}\left(\frac{b}{r_+}\right)^{2\gamma}.
\end{equation}
It should be pointed out here that since it is supposed that the pressure is an independent thermodynamic value the mass of the black hole should be treated now as the enthalpy function $M=H$ \cite{Kastor_CQG09}, in contrast to the standard framework where it is identified with the internal energy $M=E_{int}$. Taking into account the relations for the pressure (\ref{press}) and the mass (\ref{mass_bh}) we derive thermodynamic volume of the black hole by virtue of the well-known relation, namely we write:
\begin{equation}\label{TD_volume}
V=\left(\frac{\partial H}{\partial P}\right)_S=\left(\frac{\partial M}{\partial P}\right)_S=\frac{\omega_{n-1}(1+\alpha^2)}{n-\alpha^2}b^{(n-1)\gamma}r_+^{(n-1)(1-\gamma)+1}.
\end{equation} 
It is worth noting that the obtained relation completely coincides with the volume obtained for other dilaton black holes \cite{Stetsko_EPJC19, Stetsko_EYM20}.

In addition we suppose that the parameter $q$ which defines the gauge potential and field might be varied. We introduce the Yang-Mills charge in the following way:
\begin{equation}\label{YM_charge}
Q=\frac{1}{4\pi\sqrt{(n-1)(n-2)}}\int_{\Sigma}d^{n-1}\chi J(\Omega)\left(Tr(F^{(a)}_{\mu\nu}F^{(a)}_{\mu\nu})\right)^{\frac{p}{2}}=\frac{\omega_{n-1}}{4\pi}((n-1)(n-2))^{\frac{p-1}{2}}q^p.
\end{equation} 
The latter relation is a generalization of the corresponding relation for Yang-Mills charge in case of the linear gauge field theory \cite{Corichi_PRD00}. The integral in the written above relation is taken over a sphere which encloses the black hole and $J(\Omega)$ denotes the Jacobian for the spherical variables. Since we have introduced the Yang-Mills charge (\ref{YM_charge}) we derive corresponding conjugate value:
\begin{equation}
U=\left(\frac{\partial M}{\partial Q}\right)_{S,P}.
\end{equation}
The additional thermodynamic values we have introduced in this section allow us to write the so called extended first law:
\begin{equation}\label{ext_first_law}
dM=TdS+VdP+UdQ.
\end{equation}
In the framework of the extended thermodynamics we can also derive the so called Smarr relation which establish some relation between the mass $M$ of the black hole, treated as a thermodynamic function and all the mentioned thermodynamic variables. In our case we write:
\begin{equation}\label{smarr_gen}
(n+\al^2-2)M=(n-1)TS+2(\al^2-1)VP+(2p-\al^2-1)UQ.
\end{equation} 
In the linear field case ($p=1$) the latter relation completely coincides with the relation obtained in earlier work \cite{Stetsko_EYM20}. 

In the framework of the extended thermodynamics the important role plays the so called equation of state which is supposed to be an analog of thermal equation of state in ordinary condensed matter thermodynamics. In our case we obtain:
\begin{equation}\label{eos_1}
P=\frac{(n-1)}{4(1+\al^2)}\frac{T}{r_+}-\frac{(n-1)}{16\pi}\left(\frac{n-2}{1-\al^2}b^{-2\g}r^{2(\g-1)}_{+}+\frac{p(n-1)^{p-1}(n-2)^p}{\al^2-p}q^{2p}b^{-2(2p+1)\g}r^{2\g+4p(\g-1)}_{+}\right).
\end{equation}
This equation of state is analogous to the well known Van der Waals equation of state \cite{Kubiznak_CQG17}. To develop this analogy instead of ``geometrical'' quantities such as the pressure $P$ and the temperature we introduce their ``physical'' counterparts, by means of the relations: 
\begin{equation}
[P]=\frac{\hbar c}{l^{n-1}_{Pl}}P, \quad [T]=\frac{\hbar c}{k}T,
\end{equation}
where $l_{Pl}$ is the Planck length in $n$--dimensional space and $k$ is the Boltzmann constant. Since the ``physical'' quantities are just proportional to the ``geometrical'' ones we keep the latter and now can treat them as physical values. The equation (\ref{eos_1}) might be rewritten in the form:
\begin{equation}\label{eos_2}
P=\frac{T}{v}-\frac{(n-1)}{16\pi}\left(\frac{n-2}{1-\al^2}b^{-2\g}(\k v)^{2(\g-1)}+\frac{p(n-1)^{p-1}(n-2)^p}{\al^2-p}q^{2p}b^{-2(2p+1)\g}(\k v)^{2\g+4p(\g-1)}\right),
\end{equation}
where $v=4(1+\al^2)r_+/(n-1)$ is the new ``volume'' and the parameter $\k=(n-1)/(4(1+\al^2))$. The rewritten equation of state (\ref{eos_2}) might be studied similarly as the standard Van der Waals equation. The most important point for us here is the existence and the character of phase transitions for the system with the equation of state (\ref{eos_2}). As it is shown the phase transition might take place between the so-called small and large black holes \cite{Kubiznak_CQG17}. It is known that the van der Waals system possesses an inflection point, which allows to derive critical values of the temperature, the pressure and the volume. The inflection point is defined in the following way:
\begin{equation}
\left(\frac{\partial P}{\partial v}\right)_T=0, \quad \left(\frac{\partial^2 P}{\partial v^2}\right)_T=0.
\end{equation} 
Having used these relations we obtain the critical values for the volume $v_c$, temperature $T_c$ and pressure $P_c$:
\begin{equation}\label{v_c}
v_c=\frac{4(1+\al^2)}{(n-1)}\left(\frac{p(\al^2-2p)(4p-3\al^2-1)}{(\al^2-p)}((n-1)(n-2))^{p-1}q^{2p}b^{-4p\g}\right)^{\frac{1}{4p(1-\g)-2}}.
\end{equation}
\begin{equation}\label{t_c}
T_c=\frac{(n-1)(n-2)(\al^2-2p+1)}{4\pi(1-\alpha^4)(3\alpha^2-4p+1)}\kappa^{2(\gamma-1)}b^{-2\gamma}v_c^{2\gamma-1}.
\end{equation}
 \begin{equation}\label{p_c}
P_c=\frac{(n-1)(n-2)(1+\al^2-2p)}{16\pi(1+\alpha^2)(\al^2-2p)}\kappa^{2(\gamma-1)}b^{-2\gamma}v_c^{2(\gamma-1)}.
\end{equation}
Using these relations for the critical values one can write the critical ratio in the form:
\begin{equation}\label{cr_rat}
\rho_c=\frac{P_cv_c}{T_c}=\frac{(1-\alpha^2)(3\al^2-4p+1)}{4(\alpha^2-2p)}.
\end{equation}
We point out here that if $p=1$ the written above relations for the the critical values (\ref{v_c})-(\ref{cr_rat}) are reduced to the form derived in our earlier work \cite{Stetsko_EYM20} as it has to be.

\begin{figure}
\centerline{\includegraphics[scale=0.33,clip]{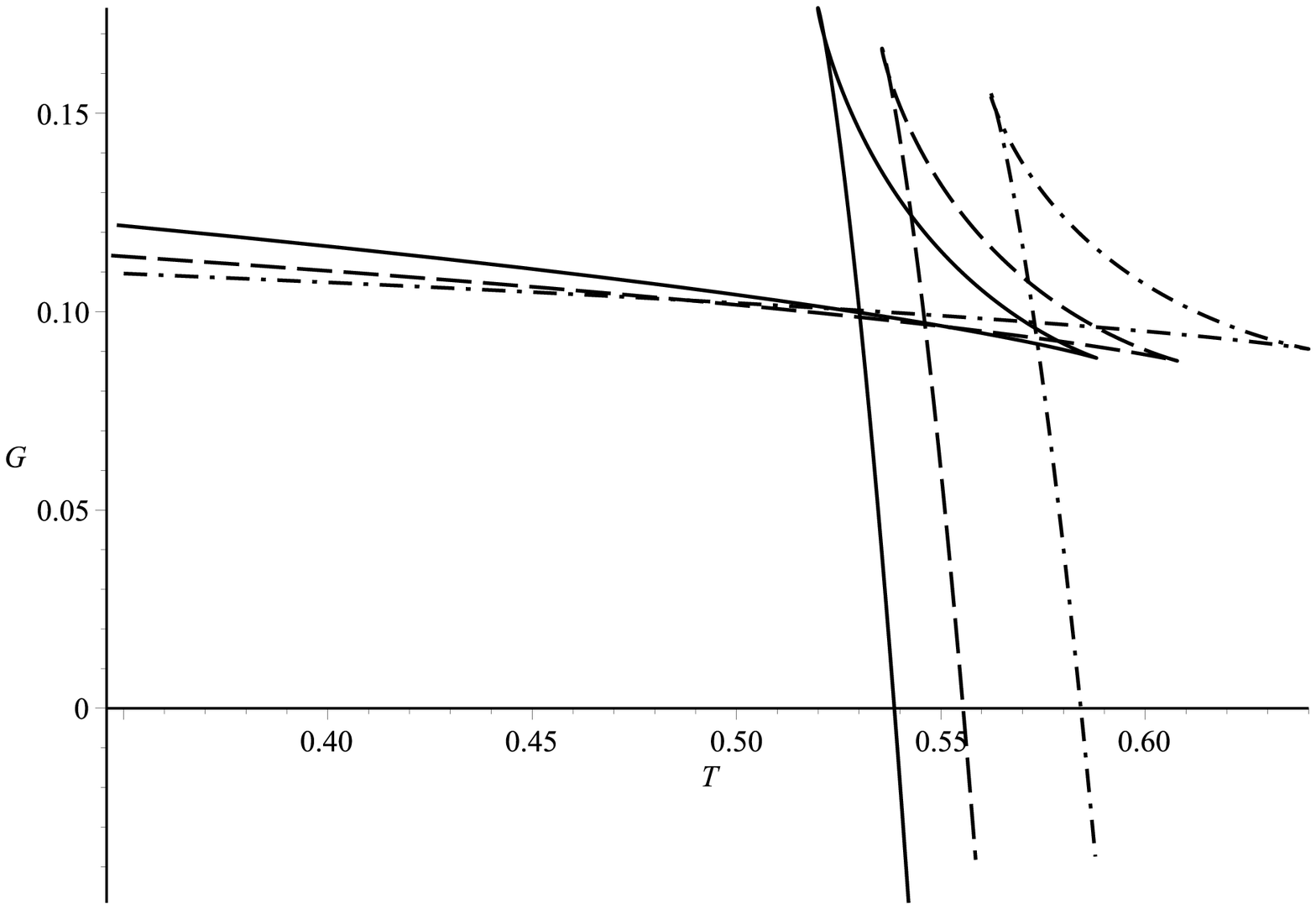}\includegraphics[scale=0.33,clip]{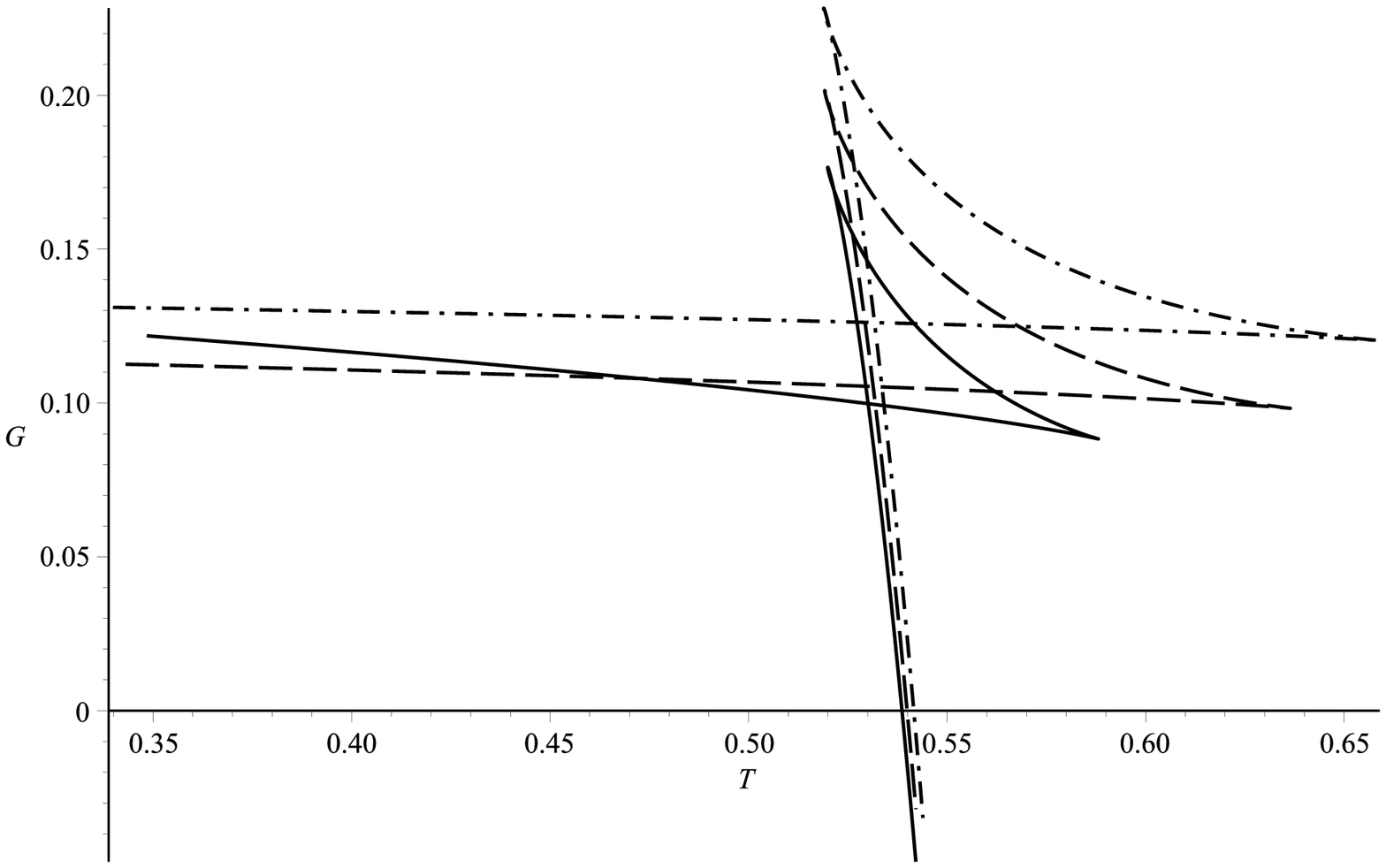}}
\caption{Gibbs free energy $G$ as a function of temperature with fixed pressure. For both graphs we have taken $n=5$, $\al=0.1$, $b=1$, $q=0.1$. For the left graph the solid, dashed and dash-dotted curves correspond to $p=2$, $p=1.8$ and $p=1.6$ respectively and the to fixed pressure equals $P_c/2$, where $P_c$ is the critical pressure for the corresponding values of the parameters. For the right graph the solid, dashed and dash-dotted lines correspond to $p=2$, $p=1.5$ and $p=1.4$ while the pressure for all of them is fixed ($P=0.276$).}\label{Gibbs_en_gr}
\end{figure}

\begin{figure}[!]
\centerline{\includegraphics[scale=0.33,clip]{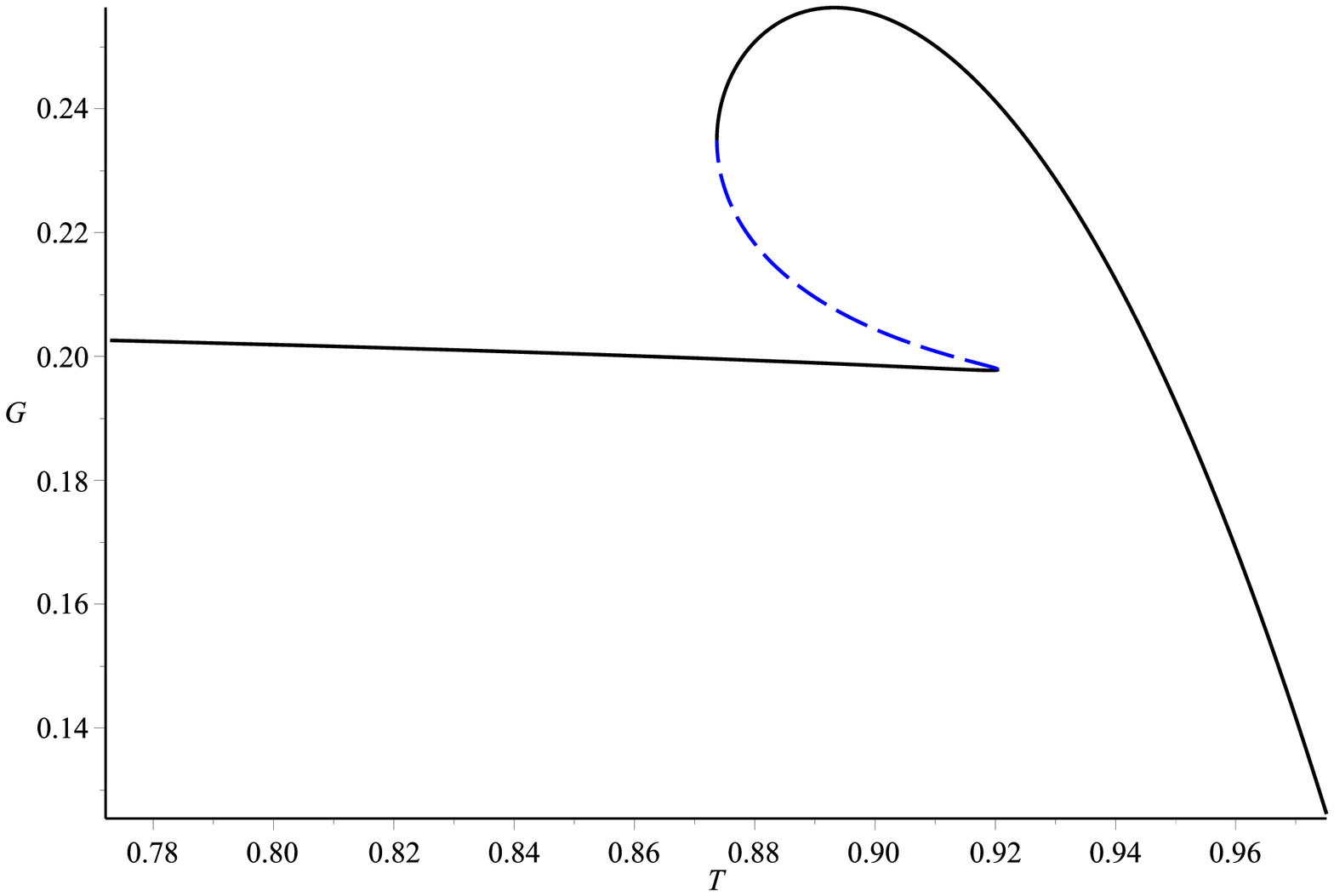}\includegraphics[scale=0.33,clip]{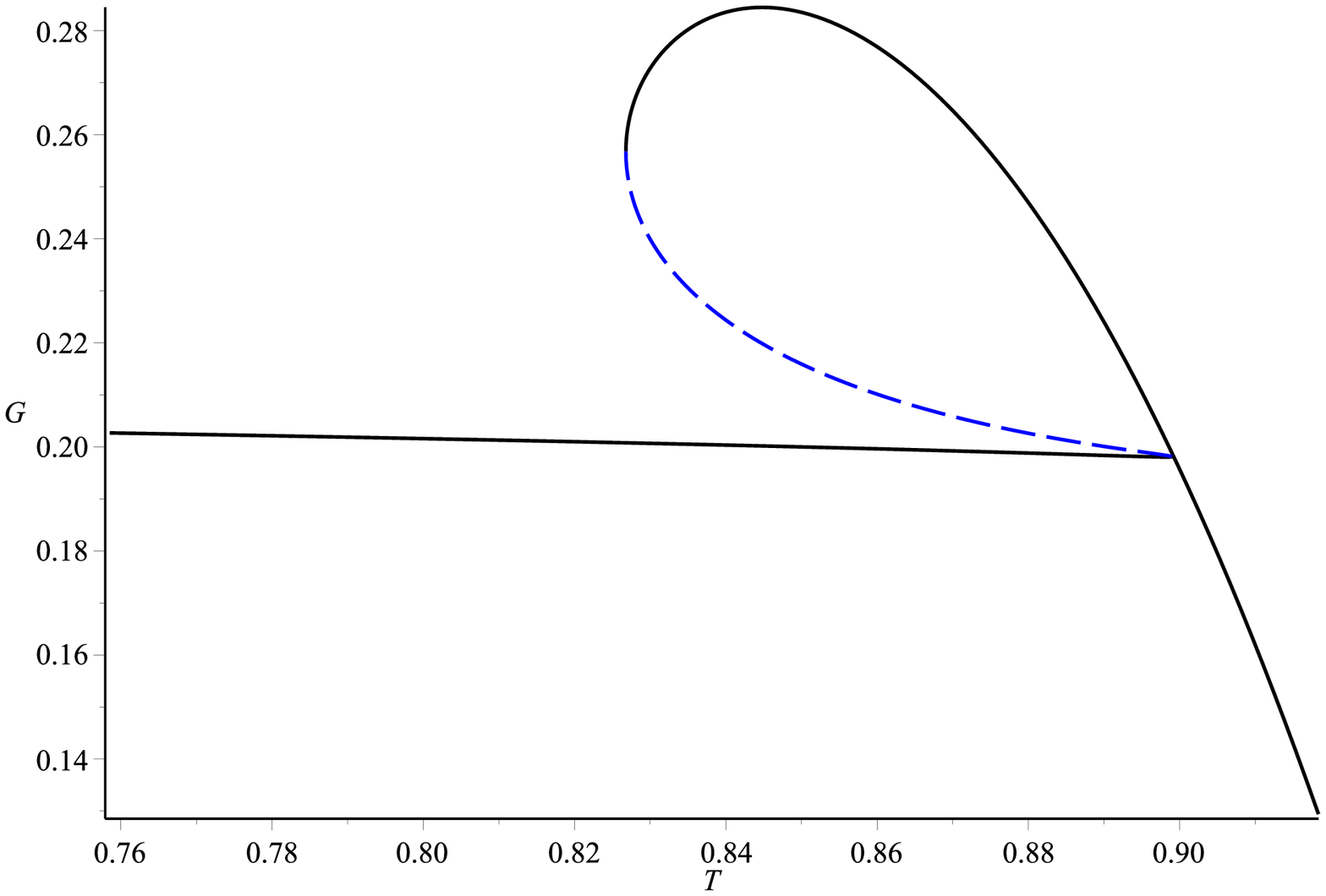}}
\caption{Gibbs free energy $G$ as a function of temperature $T$, dashed lines show instability domains. For the left graph the Gibbs free energy turns to be discontinuous and for the right graph when the specific loop is formed the discontinuity disappears.}\label{G_discont}
\end{figure}

To analyze the critical behaviour of the black hole it is convenient to introduce the Gibbs free energy instead of the enthalpy function $H=M$.  The Gibbs free energy can be written in the form:
\begin{eqnarray}
\nonumber G=\frac{\omega_{n-1}(1+\al^2)}{16\pi}b^{(n-1)\g}r^{(n-1)(1-\g)}_{+}\left(\frac{n-2}{n+\al^2-2}b^{-2\g}r^{2\g-1}_{+}+\frac{16\pi(\al^2-1)}{(n-1)(n-\al^2)}Pr_{+}\right.\\\left.+\frac{p(n-1)^{p-1}(n-2)^p(4p-3\al^2-1)}{(\al^2-p)(n+3\al^2-4p)}q^{2p}b^{-2(2p+1)\g}r^{1+2\g+4p(\g-1)}_{+}\right).
\end{eqnarray}
The function $G=G(T)$ for some fixed values of the pressure $P$ is shown graphically on the Figure [\ref{Gibbs_en_gr}]. We have chosen the values of pressure below the critical one and as a consequence for all the graphs we have the swallow-tail behaviour which is typical for systems with the phase transition of the first order. The first order phase transition was shown to take place for charged black holes under pressures below the critical and also for charged dilaton black holes, but with relatively small dilaton parameter $\al$.  If the dilaton parameter increases the domain with the zeroth order phase transition appears \cite{Dehyadegari_PRD17,Stetsko_EYM20} and the behaviour of the Gibbs free energy turns to be subtler. Namely, when $P=P_c$ the function $G=G(T)$ has a specific maximum at the critical pressure \cite{Dehyadegari_PRD17,Stetsko_EYM20}, below the critical pressure there is the domain where the zeroth order phase transition takes place. The latter fact is demonstrated on the Figure [\ref{G_discont}], namely on the left graph, when a specific loop is not closed yet we have the phase transition of the zeroth order and the following decrease of the pressure gives rise to the closing of the loop and the phase transition becomes of the first order. We also note, that on the Figure [\ref{G_discont}] the dashed parts of the curves show the domains of instability, namely at those domains the isothermal compressibility  $\k_T=-1/V(\partial V/\partial P)_T$ becomes negative.  Here we point out that similar type of behaviour takes place for Einstein-Maxwell-dilaton \cite{Dehyadegari_PRD17} as well as Einstein-Yang-Mills-dilaton \cite{Stetsko_EYM20} black holes. At the critical point the second derivatives of the Gibbs free energy are divergent and consequently such values as the mentioned above isothermal compressibility $\k_T$ and in addition heat capacity $C_p$  and volume expansion coefficient $\tilde{\al}$ will be divergent too, thus there is the phase transition of the second order. But to understand the character of the transition at the critical point we have to investigate the thermal behaviour at this point more carefully, what will be performed in the next section.

\section{Ehrenfest's equations for the second order phase transition}
It is known that for the second order phase transitions there are the so-called Ehrenfest's equations which reflect discontinuities of the heat capacity $C_P$, isothermal compressibility $\k_T$ and volume expansion coefficient $\tilde{\al}$. Namely, we can write:
\begin{eqnarray}
\left(\frac{\partial P}{\partial T}\right)_S=\frac{C_{P_2}-C_{P_1}}{VT(\tilde{\al}_2-\tilde{\al}_1)}=\frac{\Delta C_P}{VT\Delta\tilde{\al}}\label{ehrenfest_1},\\
\left(\frac{\partial P}{\partial T}\right)_V=\frac{\tilde{\al}_2-\tilde{\al}_1}{\k_{T_2}-\k_{T_1}}=\frac{\Delta\tilde{\al}}{\Delta\k_T},\label{ehrenfest_2}
\end{eqnarray}
where $C_P=T\left(\partial S/\partial T\right)_P$, $\tilde{\al}=1/V\left(\partial V/\partial T\right)_P$ and $\k_T=-1/V\left(\partial V/\partial P\right)_T$. To simplify the following calculations we rewrite the temperature $T$ (\ref{temp_nln}) now as a function of the entropy $S$, the pressure $P$ and the charge $Q$:
\begin{eqnarray}
\nonumber T=\frac{(1+\al^2)}{4\pi}\left(\frac{(n-2)}{(1-\al^2)}b^{-2\g}\left(\frac{4S}{\omega_{n-1}b^{(n-1)\g}}\right)^{\frac{2\g-1}{(n-1)(1-\g)}}+\frac{16\pi P}{(n-1)}\left(\frac{4S}{\omega_{n-1}b^{(n-1)\g}}\right)^{\frac{1}{(n-1)(1-\g)}}\right.\\\left.+\frac{16\pi^2p(n-2)}{\omega^2_{n-1}(\al^2-p)}Q^2b^{-2(2p+1)\g}\left(\frac{4S}{\omega_{n-1}b^{(n-1)\g}}\right)^{\frac{1+2\g+4p(\g-1)}{(n-1)(1-\g)}}\right).
\end{eqnarray}
Taking derivatives we can write:
\begin{eqnarray}\label{C_P}
\nonumber C_P=\frac{(n-1)S}{1+\al^2}\left(P+\frac{(n-1)(n-2)}{16\pi(1-\al^2)}b^{-2\g}\left(\frac{4S}{\omega_{n-1}b^{(n-1)\g}}\right)^{-\frac{2}{(n-1)}}+\frac{\pi p(n-1)(n-2)}{\omega^2_{n-1}(\al^2-p)}\times\right.\\\left.\nonumber Q^2b^{-2(2p+1)\g}\left(\frac{4S}{\omega_{n-1}b^{(n-1)\g}}\right)^{\frac{2\g+4p(\g-1)}{(n-1)(1-\g)}}\right)\left(P-\frac{(n-1)(n-2)}{16\pi(1+\al^2)}b^{-2\g}\left(\frac{4S}{\omega_{n-1}b^{(n-1)\g}}\right)^{-\frac{2}{(n-1)}}\right.\\\left.+\frac{\pi p(n-1)(n-2)(1+3\al^2-4p)}{\omega^2_{n-1}(\al^2-p)(1+\al^2)}Q^2b^{-2(2p+1)\g}\left(\frac{4S}{\omega_{n-1}b^{(n-1)\g}}\right)^{\frac{2\g+4p(\g-1)}{(n-1)(1-\g)}}\right)^{-1};
\end{eqnarray}
\begin{eqnarray}\label{alpha}
\nonumber\tilde{\al}=\frac{(n-1)(n+\al^2)}{4(1+\al^2)^2}\left(\frac{4S}{\omega_{n-1}b^{(n-1)\g}}\right)^{-\frac{1}{(n-1)(1-\g)}}\left(P-\frac{(n-1)(n-2)}{16\pi(1+\al^2)}b^{-2\g}\left(\frac{4S}{\omega_{n-1}b^{(n-1)\g}}\right)^{-\frac{2}{(n-1)}}\right.\\\left.+\frac{\pi p(n-1)(n-2)(1+3\al^2-4p)}{\omega^2_{n-1}(\al^2-p)(1+\al^2)}Q^2b^{-2(2p+1)\g}\left(\frac{4S}{\omega_{n-1}b^{(n-1)\g}}\right)^{\frac{2\g+4p(\g-1)}{(n-1)(1-\g)}}\right)^{-1};
\end{eqnarray}
\begin{eqnarray}\label{k_T}
\nonumber\k_{T}=\frac{n+\al^2}{1+\al^2}\left(P-\frac{(n-1)(n-2)}{16\pi(1+\al^2)}b^{-2\g}\left(\frac{4S}{\omega_{n-1}b^{(n-1)\g}}\right)^{-\frac{2}{(n-1)}}\right.\\\left.+\frac{\pi p(n-1)(n-2)(1+3\al^2-4p)}{\omega^2_{n-1}(\al^2-p)(1+\al^2)}Q^2b^{-2(2p+1)\g}\left(\frac{4S}{\omega_{n-1}b^{(n-1)\g}}\right)^{\frac{2\g+4p(\g-1)}{(n-1)(1-\g)}}\right)^{-1}.
\end{eqnarray}
Using the relation (\ref{v_c}) we can derive the entropy at the critical point (critical entropy) which can be written in the form:
\begin{equation}\label{S_c}
S_c=\frac{\omega_{n-1}}{4}b^{(n-1)\g}(\k v_c)^{(n-1)(1-\g)}.
\end{equation}
Taking into account the relations (\ref{S_c}) and (\ref{p_c}) we can check that at the critical point the heat capacity $C_P$ (\ref{C_P}), the volume expansion coefficient $\tilde{\al}$ (\ref{alpha}) and the isothermal compressibility $\k_T$ are all divergent since the denominators in the relations (\ref{C_P}), (\ref{alpha}) and (\ref{k_T}), which is common for all of them, is equal to zero, namely:
\begin{eqnarray}
\nonumber P_c-\frac{(n-1)(n-2)}{16\pi(1+\al^2)}b^{-2\g}\left(\frac{4S_c}{\omega_{n-1}b^{(n-1)\g}}\right)^{-\frac{2}{(n-1)}}\\+\frac{\pi p(n-1)(n-2)(1+3\al^2-4p)}{\omega^2_{n-1}(\al^2-p)(1+\al^2)}Q^2b^{-2(2p+1)\g}\left(\frac{4S_c}{\omega_{n-1}b^{(n-1)\g}}\right)^{\frac{2\g+4p(\g-1)}{(n-1)(1-\g)}}=0.
\end{eqnarray}
So, indeed the critical point is the point of the second order phase transition. Since, as we have pointed out above, in the relations (\ref{C_P}), (\ref{alpha}) and (\ref{k_T}) we have the divergence of the same character at the critical point it means that their relations which appear in the right hand sides of the Eherenfest's equations (\ref{ehrenfest_1}) and (\ref{ehrenfest_2}) would be finite. Namely, for the relation (\ref{ehrenfest_1}) we obtain:
\begin{equation}\label{ehr_11}
\left(\frac{\partial P}{\partial T}\right)_S=\frac{(n-1)(n-\al^2)}{4(1+\al^2)(n+\al^2)}\left(\frac{4S_c}{\omega_{n-1}b^{(n-1)\g}}\right)^{-\frac{1}{(n-1)(1-\g)}}.
\end{equation}
Then for the relation (\ref{ehrenfest_2}) we have:
\begin{equation}\label{ehr_21}
\left(\frac{\partial P}{\partial T}\right)_V=\frac{(n-1)}{4(1+\al^2)}\left(\frac{4S_c}{\omega_{n-1}b^{(n-1)\g}}\right)^{-\frac{1}{(n-1)(1-\g)}}.
\end{equation}
To describe the character of the phase transition at the critical point we also calculate Prigogine-Defay ratio, which is defined as follows:
\begin{equation}\label{PD_rat}
\Pi=\frac{\Delta C_P\Delta\kappa_T}{VT(\Delta\tilde{\al})^2}.
\end{equation} 
In our case taking into account the upper relations (\ref{ehr_11}) and (\ref{ehr_21}) we obtain:
\begin{equation}\label{PD_calc}
\Pi=\frac{n-\al^2}{n+\al^2}.
\end{equation}
From the latter relation it follows that for arbitrary $n$ and $\al\neq 0$ the Prigogine-Defay ratio $\Pi<1$. Only if $\al=0$ (dilaton coupling is absent) we obtain that $\Pi=1$. Thus, we can conclude that at the critical point we rather have the so called glass phase transition ($\Pi\neq 1$) than the ordinary phase transition of the second order for which we should have $\Pi=1$. It is worth noting that the Prigogine-Defay ratio depends on the dimension $n$ and the dilaton coupling paramater $\al$ and it does not depend on the power $p$, so the Prigogine-Defay ratio for nonlinear gauge field ($p\neq 1$) is the same as for the linear one ($p=1$). The latter fact simply means that the character of the critical point is completely defined by the dilaton field. It is also worth pointing out that the obtained relation for the Prigogine-Defay ratio (\ref{PD_calc}) has some ``universality'' similarly to the critical ratio (\ref{cr_rat}), both of them are function of universal parameters such as $n$, $\al$ and $p$ and do not contain any solution-dependent parameters such as $m$, $q$ or $b$.  Although, as we have noted above the Prigogine-Defay ratio depends just on $n$ and $\al$ and we might suppose that it would be the same if we consider other type of the gauge field or its Lagrangian, whereas the critical ratio depends on the nonlinearity parameter $p$ and certainly it would take different form for some other type of the gauge field Lagrangian. 
\section{Conclusions}
In this work we have considered Einstein-power-Yang-Mills-dilaton theory and obtained a static black hole solution in case the gauge group is $SO(n)$. Important ingredient of our action is the dilaton potential $V(\Phi)$ which is taken in the so-called Liouville form and this form allowed us to obtain the exact black hole solution which can be written in relatively simple form (\ref{W_nln}). It should be pointed out that the power-law dependence for the Yang-Mills action (\ref{action_int}) might be considered as a natural generalization of the standard linear dependence and the latter one was studied in our recent paper \cite{Stetsko_EYM20}, thus in the limit $p=1$ we recover corresponding  solution obtained for the linear field case as it has to be. As it has been shown the nonlinearity affects considerably on the behaviour of the metric function for small distances. While for the linear field we had the only horizon for arbitrary values of the parameters $m$ and $q$ for power-law nonlinear field we might have two horizon similarly as it takes place for Einstein-Maxwell-Yang-Mills-dilaton case \cite{Stetsko_EMDYM20} or even standard Reissner-Nordstrom black hole. We point out that the outer horizon is the event horizon. If the parameter $q$ (Yang-Mills charge parameter) goes up the horizons become closer and finally they merge so the black hole turns to be an extreme one. The further increase of the parameter $q$ gives rise to the appearance of a naked singularity instead of the black hole. The latter facts make the obtained solution closer to the mentioned Einstein-Maxwell-Yang-Mills-dilaton solution \cite{Stetsko_EMDYM20} than to the linear Einsten-Yang-Mills-dilaton case \cite{Stetsko_EYM20}. The metric we have derived here has several singular points, namely the horizons, infinity and the origin. To find the points of true physical singularity we examined the Kretschmann scalar, it was shown that the only point of true physical singularity is the origin, similarly as it usually takes place for black holes, whereas the horizon points are the points of a coordinate singularity which might be removed by a suitable redefinition of the coordinate system.

Thermodynamics of the obtained black hole was also studied in this work. First, we have obtained and examined the temperature of the black hole as the function of the radius of the event horizon $T=T(r_+)$.  The temperature shows some qualitative similarities with corresponding relations for the temperature obtained in our previous works \cite{Stetsko_EYM20,Stetsko_EMDYM20} where similar types of black holes, but with linear Yang-Mills field or even with the black hole obtained in the framework of Einstein-Maxwell-dilation theory \cite{Stetsko_EPJC19}. Namely, the temperature (\ref{temp_nln}) rises  when the horizon radius $r_+$ is relatively large and goes up. For small radius of the horizon the temperature goes down and finally for some intermediate values of $r_+$ the temperature shows nonmonotonous behaviour. The latter fact gives some hints about critical behaviour of the black hole.  We have also calculated the heat capacity $C_{\L,q}$ of the black holes which allows to characterize stability-instability regions. Namely for relatively large $r_+$ the heat capacity is positive and the black hole is stable. Then, due to nonmonotonous character of the temperature we have discontinuities of the heat capacity, the discontinuity points separate the stable and unstable domains. Thus, for some intermediate values of the horizon radius $r_+$ the black hole is unstable and for smaller values of $r_+$ below the second discontinuity point the heat capacity is again positive and it means that the black hole again becomes stable. It should be pointed out that the increase of the cosmological constant in absolute value gives rise to the fact that the discontinuity points become closer and the further increase of the module of $\L$ leads to disappearance of the discontinuity and the black holes turns to be stable on all this domain. We also point out here that increase of the parameter of nonlinearity $p$ leads to the shift of the discontinuity points to the right while the qualitative behaviour of the temperature and the heat capacity remains the same as for linear field $p=1$.

Using the extended thermodynamics concept, namely assuming that the cosmological constant $\L$ might be varied and identifying it with the thermodynamic pressure we derived the equation of state for the black hole (\ref{eos_1}). It is known that this equation of state might be treated as an analog of the van der Waals equation of state \cite{Kubiznak_CQG17}. The extended thermodynamics concept also allowed us to derive the Smarr relation (\ref{smarr_gen}) what is not possible to obtain if the parameter $\L$ is kept fixed and is not treated as a thermodynamic value. Using the approach developed for the van der Waals equation of state we obtained critical values for the volume $v_c$, the pressure $P_c$ and the temperature $T_c$ and calculated critical ratio (\ref{cr_rat}). The critical ratio in our case depends only on the dilaton $\al$ and nonlinearity $p$ parameters and if $p=1$ it reduced to the corresponding value obtained in our earlier work \cite{Stetsko_EYM20}. We also note that when $\al=0$ and $p=1$ the critical ratio is equal to $3/8$ what exactly coincides with corresponding value for ordinary van der Waals system in three dimensional space. We also derived the the Gibbs free energy which allowed us to describe the thermodynamic behaviour not only near the critical point but in wider domain. The analysis of the Gibbs free energy as a function of the temperature while the pressure is kept fixed shows that for the pressures below the critical one there is the range of pressures where the zeroth order phase transition takes place and with the following decrease of the pressure we have the phase transition of the first order, what is typical for ordinary van der Waals systems as well as for numerous black holes where corresponding analog of the van der Waals equation of state can be constructed. We also point out that the range of pressures where the zeroth order phase transition takes place becomes wider when the dilaton parameter $\al$ goes up and correspondingly it disappear when $\al\to 0$. The obtained van der Waals equation of state also allows to conclude that at the critical point takes the same values as for other types of the black holes \cite{Stetsko_EYM20,Stetsko_EPJC19,Majhi_PLB17}. We have also calculated Prigogine-Defay ratio $\Pi$ (\ref{PD_rat}) which might give additional information about the character of  the phase transition at the critical point. In particular for the standard second order phase transition this ratio is equal to one. If it is not equal to one the phase transition at the critical point is supposed to be of glass type \cite{Tropin_JCP12}, here we note that the ratio $\Pi$ for the glass transition is more than one, but it might happen that it is less than one \cite{Garden_JNET12} and in our case we have that $\Pi<1$ (\ref{PD_calc}) and it tends to one when the dilaton coupling parameter $\al$ goes to zero. We also point out that the Prigogine-Defay ratio is completely defined by the dimension $n$ and the parameter $\al$ and does not depend on the parameter of nonlinearity $p$ an we might assume that it takes the same value for other types of the gauge fields which are coupled to the dilaton field in similar way. 

\section{Acknowledgments}
This work was partly supported by Project FF-83F (No. 0119U002203) from the Ministry of Education and Science of Ukraine.  

\end{document}